# SARS-CoV-2 (COVID-19) by the numbers

Yinon M. Bar-On[1], Avi Flamholz[2], Rob Phillips[3,4], and Ron Milo[1*]
[1]Weizmann Intitute of Science, Rehovot 7610001, Israel [2]University of California, Berkeley, CA 94720, USA
[3]California Institute of Technology, Pasadena, CA 91125, USA [4]Chan Zuckerberg Biohub, San Francisco, CA 94158, USA
*Corresponding author: ron.milo@weizmann.ac.il.
Comments are welcome; this article is being updated on an ongoing basis at: https://bit.ly/2WOeN64

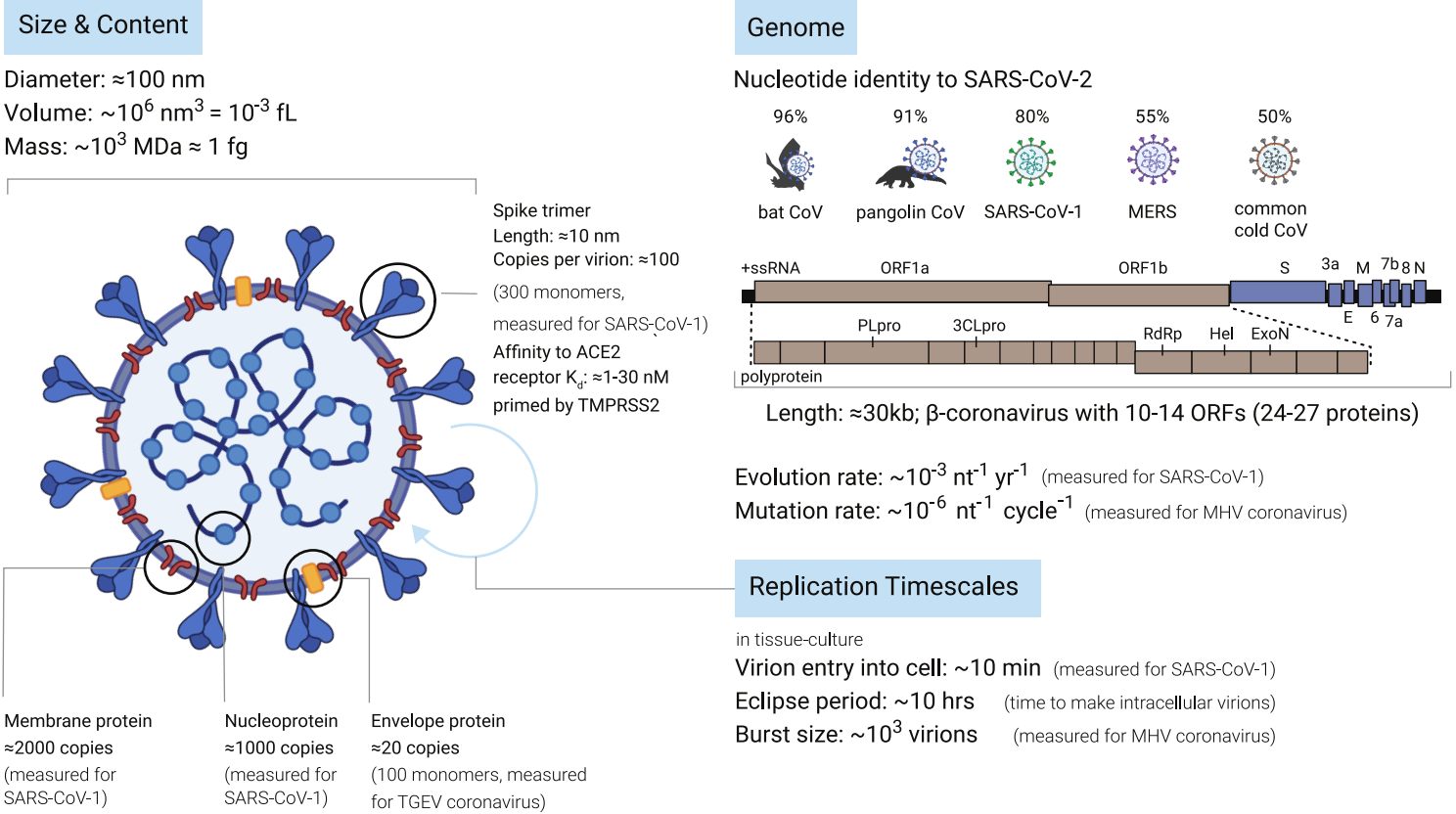

### Size & Content

Diameter: ≈100 nm
Volume: ~$10^6$ nm$^3$ = $10^{-3}$ fL
Mass: ~$10^3$ MDa ≈ 1 fg

Spike trimer
Length: ≈10 nm
Copies per virion: ≈100
(300 monomers, measured for SARS-CoV-1)
Affinity to ACE2 receptor $K_d$: ≈1-30 nM
primed by TMPRSS2

Membrane protein ≈2000 copies (measured for SARS-CoV-1)
Nucleoprotein ≈1000 copies (measured for SARS-CoV-1)
Envelope protein ≈20 copies (100 monomers, measured for TGEV coronavirus)

### Genome

Nucleotide identity to SARS-CoV-2

| 96% | 91% | 80% | 55% | 50% |
|---|---|---|---|---|
| bat CoV | pangolin CoV | SARS-CoV-1 | MERS | common cold CoV |

Length: ≈30kb; β-coronavirus with 10-14 ORFs (24-27 proteins)

Evolution rate: ~$10^{-3}$ nt$^{-1}$ yr$^{-1}$ (measured for SARS-CoV-1)
Mutation rate: ~$10^{-6}$ nt$^{-1}$ cycle$^{-1}$ (measured for MHV coronavirus)

### Replication Timescales

in tissue-culture
Virion entry into cell: ~10 min (measured for SARS-CoV-1)
Eclipse period: ~10 hrs (time to make intracellular virions)
Burst size: ~$10^3$ virions (measured for MHV coronavirus)

### Host Cells

(tentative list; number of cells per person)
Type I & II pneumocytes (~$10^{11}$ cells)
Alveolar macrophage (~$10^{10}$ cells)
Mucous cell in nasal cavity (~$10^9$ cells)
Host cell volume: ~$10^3$ μm$^3$ = $10^3$ fL

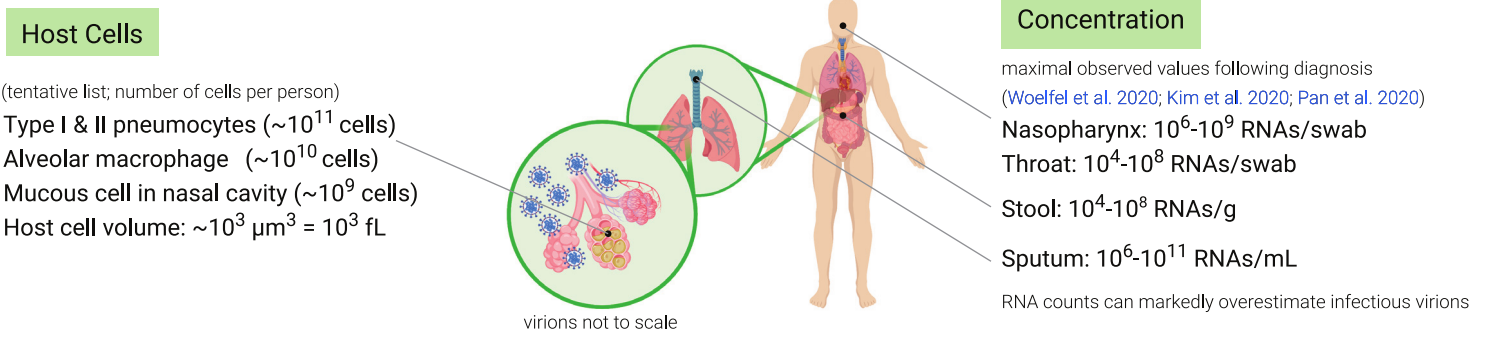

virions not to scale

### Concentration

maximal observed values following diagnosis
(Woelfel et al. 2020; Kim et al. 2020; Pan et al. 2020)
Nasopharynx: $10^6$-$10^9$ RNAs/swab
Throat: $10^4$-$10^8$ RNAs/swab
Stool: $10^4$-$10^8$ RNAs/g
Sputum: $10^6$-$10^{11}$ RNAs/mL

RNA counts can markedly overestimate infectious virions

### Antibody Response - Seroconversion

Antibodies appear in blood after: ≈10-20 days
Maintenance of antibody response:
≈2-3 years (measured for SARS-CoV-1)

### Virus Environmental Stability

Relevance to personal safety unclear

|  | half-life | time to decay 1000-fold |
|---|---|---|
| Aerosols: | ≈1 hr | ≈4-24 hr |
| Surfaces: e.g. plastic, cardboard and metals | ≈1-7 hr | ≈4-96 hr |

(van Doremalen et al. 2020)

Based on quantifying infectious virions. Tested at 21-23°C and 40-65% relative humidity. Numbers will vary between conditions and surface types (Otter et al. 2016).
Viral RNA observed on surfaces even after a few weeks (Moriarty et al. 2020).

### "Characteristic" Infection Progression in a Single Patient

Basic reproductive number $R_0$: typically 2-4
Varies further across space and time (Li et al. 2020; Park et al. 2020)
(number of new cases directly generated from a single case)

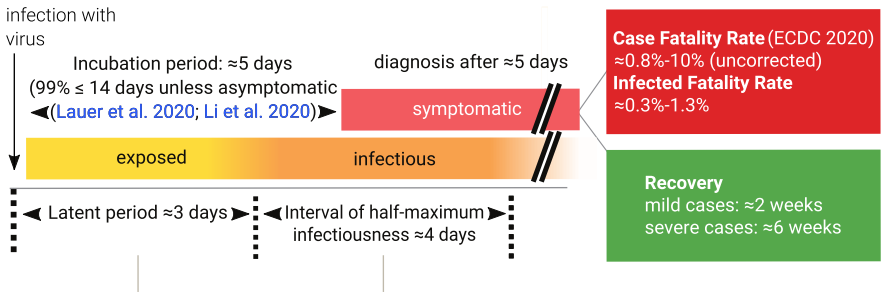

infection with virus
Incubation period: ≈5 days
(99% ≤ 14 days unless asymptomatic
◄(Lauer et al. 2020; Li et al. 2020)►
diagnosis after ≈5 days
exposed | infectious | symptomatic
◄Latent period ≈3 days► ◄Interval of half-maximum infectiousness ≈4 days►

**Case Fatality Rate** (ECDC 2020)
≈0.8%-10% (uncorrected)
**Infected Fatality Rate**
≈0.3%-1.3%

**Recovery**
mild cases: ≈2 weeks
severe cases: ≈6 weeks

Inter-individual variability is substantial and not well characterized. The estimates are parameter fits for population median in China and do not describe this variability (Li et al. 2020; He et al. 2020).

Note the difference in notation between the symbol ≈, which indicates "approximately" and connotes accuracy to within a factor 2, and the symbol ~, which indicates "order of magnitude" or accuracy to within a factor of 10.

# SARS-CoV-2 (COVID-19) by the numbers

Yinon M. Bar-On[1], Avi Flamholz[2], Rob Phillips[3,4], and Ron Milo[1]

[1]Weizmann Institute of Science, Rehovot 7610001, Israel
[2]University of California, Berkeley, CA 94720, USA
[3]California Institute of Technology, Pasadena, CA 91125, USA
[4]Chan Zuckerberg Biohub, 499 Illinois Street, SF CA 94158, USA

## Abstract

The current SARS-CoV-2 pandemic is a harsh reminder of the fact that, whether in a single human host or a wave of infection across continents, viral dynamics is often a story about the numbers. In this snapshot, our aim is to provide a one-stop, curated graphical source for the key numbers that help us understand the virus driving our current global crisis. The discussion is framed around two broad themes: 1) the biology of the virus itself and 2) the characteristics of the infection of a single human host. Our one-page summary provides the key numbers pertaining to SARS-CoV-2, based mostly on peer-reviewed literature. The numbers reported in summary format are substantiated by the annotated references below. Readers are urged to remember that much uncertainty remains and knowledge of this pandemic and the virus driving it is rapidly evolving. In the paragraphs below we provide "back of the envelope" calculations that exemplify the insights that can be gained from knowing some key numbers and using quantitative logic. These calculations serve to improve our intuition through sanity checks, but do not replace detailed epidemiological analysis.

## 1. How long does it take a single infected person to yield one million infected people?

If everybody continued to behave as usual, how long would it take the pandemic to spread from one person to a million infected victims? The basic reproduction number, $R_0$, suggests each infection directly generates 2-4 more infections in the absence of countermeasures like social distancing. Once a person is infected, it takes a period of time known as the latent period before they are able to transmit the virus. The current best-estimate of the median latent time is ≈3 days followed by a ≈4 day infectious period (Li et al. 2020, He et al. 2020). The exact durations vary among people, and some are infectious for much longer. Using $R_0$=4, the number of cases will quadruple every ≈7 days or double every ≈3 days. 1000-fold growth (going from one case to $10^3$) requires 10 doublings since $2^{10} ≈ 10^3$; 3 days × 10 doublings = 30 days, or about one month. So we expect ≈1000x growth in one month, million-fold ($10^6$) in two months, and a billion fold ($10^9$) in three months. Even though this calculation is highly simplified, ignoring the effects of "super-spreaders" and incomplete testing, it emphasizes the fact that viruses can spread at a bewildering pace when no countermeasures are taken. This illustrates why it is crucial to limit the spread of the virus by social distancing measures. For fuller discussion of the meaning of $R_0$, the latent and infectious periods, as well as various caveats, see the "Definitions" section.

## 2. What is the effect of social distancing?

A highly simplified quantitative example helps clarify the need for social distancing. Suppose that you are infected and you encounter 50 people over the course of a day of working, commuting, socializing and running errands. To make the numbers round, let's further suppose that you have a 2% chance of transmitting the virus in each of these encounters, so that you are likely to infect 1 new person each day. If you are infectious for 4 days, then you will infect 4 others on average, which is on the high end of the $R_0$ values for SARS-CoV-2 in the absence of social distancing. If you instead see 5 people each day (preferably fewer) because of social distancing, then you will infect 0.1 people per day, or 0.4 people before you become less infectious. The desired effect of social distancing is to make each current infection produce < 1 new infections. An effective reproduction number ($R_e$) smaller than 1 will ensure the number of infections eventually dwindles. It is critically important to quickly achieve $R_e$ < 1, which is substantially more achievable than pushing $R_e$ to near zero through public health measures.

## 3. Why is the quarantine period two weeks?

The period of time from infection to symptoms is termed the incubation period. The median SARS-CoV-2 incubation period is estimated to be roughly 5 days (Lauer et al. 2020). Yet there is much person-to-person variation. Approximately 99% of those showing symptoms will show them before day 14, which explains the two week confinement period. Importantly, this analysis neglects infected people who never show symptoms. Since asymptomatic people are not usually tested, it is still not clear how many such cases there are or how long asymptomatic people remain infectious for.

## 4. How do N95 masks block SARS-CoV-2?

N95 masks are designed to remove more than 95% of all particles that are at least 0.3 microns (μm) in diameter (NIOSH 42 CFR Part 84). In fact, measurements of the particle filtration efficiency of N95 masks show that they are capable of filtering ≈99.8% of particles with a diameter of ~0.1 μm (Regnasamy et al. 2017). SARS-CoV-2 is an enveloped virus ~0.1 μm in diameter, so N95 masks are capable of filtering most free virions, but they do more than that. How so? Viruses are often transmitted through respiratory droplets produced by coughing and sneezing. Respiratory droplets are usually divided into two size bins, large droplets (> 5 μm in diameter) that fall rapidly to the ground and are thus transmitted only over short distances, and small droplets (≤ 5 μm in diameter). Small droplets can evaporate into "droplet nuclei," remain suspended in air for significant periods of time and could be inhaled. Some viruses, such as measles, can be transmitted by droplet nuclei (Tellier et al. 2019). At present there is no direct evidence showing SARS-CoV-2 transmission by droplet nuclei. Rather, larger droplets are believed to be the main vector of SARS-CoV-2 transmission, usually by settling onto surfaces that are touched and transported by hands onto mucosal membranes such as the eyes, nose and mouth (CDC 2020). The characteristic diameter of large droplets produced by sneezing is ~100 μm (Han J. R. Soc. Interface 2013), while the diameter of droplet nuclei produced by coughing is on the order of ~1 μm (Yang et al 2007). Therefore, N95 masks likely protect against several modes of viral transmission.

## 5. How similar is SARS-CoV-2 to the common cold and flu viruses?

SARS-CoV-2 is a beta-coronavirus whose genome is a single ≈30 kb strand of RNA. The flu is caused by an entirely different family of RNA viruses called influenza viruses. Flu viruses have smaller genomes (≈14 kb) encoded in 8 distinct strands of RNA, and they infect human cells in a different manner than coronaviruses. The "common cold" is caused by a variety of viruses, including some coronaviruses and rhinoviruses. Cold-causing coronaviruses (e.g. OC43 and 229E strains) are quite similar to SARS-CoV-2 in genome length (within 10%) and gene content, but different from SARS-CoV-2 in sequence (≈50% nucleotide identity) and infection severity. One interesting facet of coronaviruses is that they have the largest genomes of any known RNA viruses (≈30 kb). These large genomes led researchers to suspect the presence of a "proofreading mechanism" to reduce the mutation rate and stabilize the genome. Indeed, coronaviruses have a proofreading exonuclease called ExoN, which explains their very low mutation rates (~$10^{-6}$ per site per cycle) in comparison to influenza (≈$3×10^{-5}$ per site per cycle (Sanjuan et al. 2010)). This relatively low mutation rate will be of interest for future studies predicting the speed with which coronaviruses can evade our immunization efforts.

## 6. How much is known about the SARS-CoV-2 genome and proteome?

SARS-CoV-2 has a single-stranded positive-sense RNA genome that codes for 10 genes ultimately producing 26 proteins according to an NCBI annotation (NC_045512). How is it that 10 genes code for >20 proteins? One long gene, orf1ab, encodes a polyprotein that is cleaved into 16 proteins by proteases that are themselves part of the polyprotein. In addition to proteases, the polyprotein encodes an RNA polymerase and associated factors to copy the genome, a proofreading exonuclease, and several other non-structural proteins. The remaining genes predominantly code for structural components of the virus: (i) the spike protein which binds the cognate receptor on a human or animal cell; (ii) a nucleoprotein that packages the genome; and (iii) two membrane-bound proteins. Though much current work is centered on understanding the role of "accessory" proteins in the viral life cycle, we estimate that it is currently possible to ascribe clear biochemical or structural functions to only about half of SARS-CoV-2 gene products.

## 7. What can we learn from the mutation rate of the virus?

Studying viral evolution, researchers commonly use two measures describing the rate of genomic change. The first is the evolutionary rate, which is defined as the average number of substitutions that become fixed per year in strains of the virus, given in units of mutations per site per year. The second is the mutation rate, which is the number of substitutions per site per replication cycle. How can we relate these two values? Consider a single site at the end of a year. The only measurement of a mutation rate in a β-coronavirus suggests that this site will accumulate ~$10^{-6}$ mutations in each round of replication. Each round of replication cycle takes ~10 hours, and so there are $10^3$ cycles/year. Multiplying the mutation rate by the number of replications, and neglecting the potential effects of evolutionary selection and drift, we arrive at $10^{-3}$ mutations per site per year, consistent with the evolutionary rate inferred from sequenced coronavirus genomes. As our estimate is consistent with the measured rate, we infer that the virus undergoes near-continuous replication in the wild, constantly generating new mutations that accumulate over the course of the year. Using our knowledge of the mutation rate, we can also draw inferences about single infections. For example, since the mutation rate is ~$10^{-6}$ mutations/site/cycle and an mL of sputum might contain upwards of $10^7$ viral RNAs, we infer that every site is mutated more than once in such samples.

## 8. How stable and infectious is the virion on surfaces?

SARS-CoV-2 RNA has been detected on various surfaces several weeks after they were last touched (Moriarty et al. 2020). In the definitions we clarify the difference between detecting viral RNA and active virus. The probability of human infection from such exposure is not yet characterized as experiments to make this determination are very challenging. Nevertheless, caution and protective measures must be taken. We estimate that during the infectious period an undiagnosed infectious person touches surfaces tens of times. These surfaces will subsequently be touched by hundreds of other people. From the basic reproduction number $R_0$ ≈2-4 we can infer that not everyone touching those surfaces will be infected. More detailed bounds on the risk of infection from touching surfaces urgently awaits study.

# Glossary
## Clinical Measures
**Incubation period**: time between exposure and symptoms.
**Seroconversion**: time between exposure to virus and detectable antibody response.

## Epidemiological Inferences
**$R_0$**: the average number of cases directly generated by an individual infection.
**Latent period:** time between exposure and becoming infective.
**Infectious period:** time for which an individual is infective.
**Interval of half-maximum infectiousness:** the time interval during which the probability of viral transmission is higher than half of the peak infectiousness. This interval is similar to the infectious period, but applies also in cases where the probability of infection is not uniform in time.

## Viral Species
**SARS-CoV-2**: Severe acute respiratory syndrome coronavirus 2. A β-coronavirus causing the present COVID-19 outbreak.
**SARS-CoV-1**: β-coronavirus that caused the 2002 SARS outbreak in China.
**MERS**: a β-coronavirus that caused the Middle East Respiratory Syndrome outbreak beginning in Jordan in 2012.
**MHV**: Murine herpes virus, a model β-coronavirus on which much laboratory research has been conducted.
**TGEV**: Transmissible gastroenteritis virus, a model α-coronavirus which infects pigs.
**229E and OC43**: two strains of coronavirus (α- and β- respectively) that are cause a fraction of common colds.

## Viral Life-Cycle
**Eclipse period:** time between viral entry and appearance of intracellular virions.
**Latent period (cellular level):** time between viral entry and appearance of extracellular virions. Not to be confused with the epidemiological latent period described below.
**Burst size**: the number of virions produced from infection of a single cell. More appropriately called "per-cell viral yield" for non-lytic viruses like SARS-CoV-2.
**Virion**: a viral particle.
**Polyprotein**: a long protein that is proteolytically cleaved into a number of distinct proteins. Distinct from a polypeptide, which is a linear chain of amino acids making up a protein.

## Human Biology
**Alveolar Macrophage:** immune cells found in the lung that engulf foreign material like dust and microbes ("professional phagocytes")
**Pneumocytes:** the non-immune cells in the lung.
**ACE2:** Angiotensin-converting enzyme 2, the mammalian cell surface receptor that SARS-CoV-2 binds.
**TMPRSS2:** Transmembrane protease, serine 2, a mammalian membrane-bound serine protease that cleaves the viral spike trimer after it binds ACE2, revealing a fusion peptide that participates in membrane fusion which enables subsequent injection of viral DNA into the host cytoplasm.
**Nasopharynx:** the space above the soft palate at the back of the nose which connects the nose to the mouth.

## Notation
Note the difference in notation between the symbol ≈, which indicates "approximately" and connotes accuracy to within a factor 2, and the symbol ~, which indicates "order of magnitude" or accuracy to within a factor of 10.

# More on definitions and measurement methods
## What are the meanings of $R_0$, "latent period" and "infectious period"?
The basic reproduction number, $R_0$, estimates the average number of new infections directly generated by a single infectious person. The 0 subscript connotes that this refers to early stages of an epidemic, when everyone in the region is susceptible (i.e. there is no immunity) and no counter-measures have been taken. As geography and culture affect how many people we encounter daily, how much we touch them and share food with them, estimates of $R_0$ can vary between locales. Moreover, because $R_0$ is defined in the absence of countermeasures and immunity, we are usually only able to assess the effective R ($R_e$). At the beginning of an epidemic, before any countermeasures, $R_e ≈ R_0$. Several days pass before a newly-infected person becomes infectious themselves. This "latent period" is typically followed by several days of infectivity called the "infectious period." It is important to understand that reported values for all these parameters are population averages inferred from epidemiological models fit to counts of infected, symptomatic, and dying patients. Because testing is always incomplete and model fitting is imperfect, and data will vary between different locations, there is substantial uncertainty associated with reported values. Moreover, these median or average best-fit values do not describe person-to-person variation. For example, viral RNA was detectable in patients with moderate symptoms for > 1 week after the onset of symptoms, and more than 2 weeks in patients with severe symptoms (ECDC 2020). Though detectable RNA is not the same as active virus, this evidence calls for caution in using uncertain, average parameters to describe a pandemic. Why aren't detailed distributions of these parameters across people published? Direct measurement of latent and infectious periods at the individual level is extremely challenging, as accurately identifying the precise time of infection is usually very difficult.

## What is the difference between measurements of viral RNA and infectious viruses?
Diagnosis and quantification of viruses utilizes several different methodologies. One common approach is to quantify the amount of viral RNA in an environmental (e.g. surface) or clinical (e.g. sputum) sample via quantitative reverse-transcription polymerase chain reaction (RT-qPCR). This method measures the number of copies of viral RNA in a sample. The presence of viral RNA does not necessarily imply the presence of infectious virions. Virions could be defective (e.g. by mutation) or might have been deactivated by environmental conditions. To assess the concentration of infectious viruses, researchers typically measure the "50% tissue-culture infectious dose" ($TCID_{50}$). Measuring $TCID_{50}$ involves infecting replicate cultures of susceptible cells with dilutions of the virus and noting the dilution at which half the replicate dishes become infected. Viral counts reported by $TCID_{50}$ tend to be much lower than RT-qPCR measurements, which could be one reason why studies relying on RNA measurements (Moriarty et al. 2020) report the persistence of viral RNA on surfaces for much longer times than studies relying on $TCID_{50}$ (van Doremalen et al. 2020). It is important to keep this caveat in mind when interpreting data about viral loads, for example a report measuring viral RNA in patient stool samples for several days after recovery (We et al. 2020). Nevertheless, for many viruses even a small dose of virions can lead to infection. For the common cold, for example, ~0.1 $TCID_{50}$ are sufficient to infect half of the people exposed (Couch et al. 1966).

## What is the difference between the case fatality rate and the infection fatality rate?
Global statistics on new infections and fatalities are pouring in from many countries, providing somewhat different views on the severity and progression of the pandemic. Assessing the severity of the pandemic is critical for policy making and thus much effort has been put into quantification. The most common measure for the severity of a disease is the fatality rate. One commonly reported measure is the case fatality rate (CFR), which is the proportion of fatalities out of total diagnosed cases. The CFR reported in different countries varies significantly, from 0.3% to about 10%. Several key factors affect the CFR. First, demographic parameters and practices associated with increased or decreased risk differ greatly across societies. For example, the prevalence of smoking, the average age of the population, and the capacity of the healthcare system. Indeed, the majority of people dying from SARS-CoV-2 have a preexisting condition such as cardiovascular disease or smoking (China CDC 2020). There is also potential for bias in estimating the CFR. For example, a tendency to identify more severe cases (selection bias) will tend to overestimate the CFR. On the other hand, there is usually a delay between the onset of symptoms and death, which can lead to an underestimate of the CFR early in the progression of an epidemic. Even when correcting for these factors, the CFR does not give a complete picture as many cases with mild or no symptoms are not tested. Thus, the CFR will tend to overestimate the rate of fatalities per infected person, termed the infection fatality rate (IFR). Estimating the total number of infected people is usually accomplished by testing a random sample for anti-viral antibodies, whose presence indicates that the patient was previously infected. As of writing, such assays are not widely available, and so researchers resort to surrogate datasets generated bytesting of foreign citizens returning home from infected countries (Verity et al. 2020), or epidemiological models estimating the number of undocumented cases (Li et al. 2020). These methods provide a first glimpse of the true severity of the disease.

## What is the burst size and the replication time of the virus?
Two important characteristics of the viral life cycle are the time it takes them to produce new infectious progeny, and the number of progeny each infected cell produces. The yield of new virions per infected cell is more clearly defined in lytic viruses, such as those infecting bacteria (bacteriophages), as viruses replicate within the cell and subsequently lyse the cell to release a "burst" of progeny. This measure is usually termed "burst size." SARS-CoV-2 does not release its progeny by lysing the cell, but rather by continuous budding (Park et al. 2020). Even though there is no "burst", we can still estimate the average number of virions produced by a single infected cell. Measuring the time to complete a replication cycle or the burst size *in vivo* is very challenging, and thus researchers usually resort to measuring these values in tissue-culture. There are various ways to estimate these quantities, but a common and simple one is using "one-step" growth dynamics. The key principle of this method is to ensure that only a single replication cycle occurs. This is typically achieved by infecting the cells with a large number of virions, such that every cell gets infected, thus leaving no opportunity for secondary infections. Assuming entry of the virus to the cells is rapid (we estimate 10 minutes for SARS-CoV-2), the time it takes to produce progeny can be estimated by quantifying the lag between inoculation and the appearance of new intracellular virions, also known as the "eclipse period". This eclipse period does not account for the time it takes to release new virions from the cell. The time from cell entry until the appearance of the first extracellular viruses, known as the "latent period" (not to be confused with the epidemiological latent period, see Glossary), estimates the duration of the full replication cycle. The burst size can be estimated by waiting until virion production saturates, and then dividing the total virion yield by the number of cells infected. While both the time to complete a replication cycle and the burst size may vary significantly in an animal host due to factors including the type of cell infected or the action of the immune system, these numbers provide us with an approximate quantitative view of the viral life-cycle at the cellular level.

# References and excerpts

Note that for about 10 out of 45 parameters, the literature values are from other coronaviruses. We await corresponding measurements for SARS-CoV-2.

## Size & Content

Diameter: (Zhu et al. 2020) - *"Electron micrographs of negative-stained 2019-nCoV particles were generally spherical with some pleomorphism (Figure 3). Diameter varied from about 60 to 140 nm."*
Volume- Using diameter and assuming the virus is a sphere
Mass: Using the volume and a density of ~ 1 g per mL
Number of surface spikes trimers: (Neuman et al. 2011) - *"Our model predicts ~90 spikes per particle."*
Length of surface spikes trimers: (Zhu et al. 2020) - *" Virus particles had quite distinctive spikes, about 9 to 12 nm, and gave virions the appearance of a solar corona."*
Receptor binding affinity ($K_d$): (Walls et al. 2020) - Walls et al. reports $K_d$ of ≈1 nM for the binding domain in Table 1 using Biolayer interferometry with $k_{on}$ of ≈1.5×10$^5$ M$^{-1}$ s$^{-1}$ and $k_{off}$ of ≈1.6×10$^{-4}$ s$^{-1}$.
(Wrapp et al. 2020) - Wrapp et al. reports $K_d$ of ≈15 nM for the spike (Fig.3) and ≈35 nM for the binding domain (Fig.4) using surface plasmon resonance with $k_{on}$ of ≈1.9×10$^5$ M$^{-1}$ s$^{-1}$ and $k_{off}$ of ≈2.8×10$^{-3}$ s$^{-1}$ for the spike and $k_{on}$ of ≈1.4×10$^5$ M$^{-1}$ s$^{-1}$ and $k_{off}$ of ≈4.7×10$^{-3}$ s$^{-1}$ for the binding domain. The main disagreement between the studies seems to be on the $k_{off}$.
Membrane (M; 222 aa): (Neuman et al. 2011) - *"Using the M spacing data for each virus (Fig 6C), this would give ~1100 M2 molecules per average SARS-CoV, MHV and FCoV particle"*
Envelope (E; 75 aa): (Godet et al. 1992) - *"Based on the estimated molar ratio and assuming that coronavirions bear 100 (Roseto et al., 1982) to 200 spikes, each composed of 3 S molecules (Delmas and Laude, 1990) it can be inferred that approximately 15- 30 copies of ORF4 protein are incorporated into TGEV virions (Purdue strain)."*
Nucleoprotein (364 aa): (Neuman et al. 2011) - *"Estimated ratios of M to N protein in purified coronaviruses range from about 3M:1N (Cavanagh, 1983, Escors et al., 2001b) to 1M:1N (Hogue and Brian, 1986, Liu and Inglis, 1991), giving 730–2200 N molecules per virion."*

## Genome

Type: (ViralZone) +ssRNA *"Monopartite, linear ssRNA(+) genome"*
Genome length: (Wu et al. 2020) - Figure 2
Number of genes: (Wu et al. 2020) - *"SARS-CoV-2 genome has 10 open reading frames (Fig. 2A)."* or (Wu et al. 2020) - *"The 2019-nCoV genome was annotated to possess 14 ORFs encoding 27 proteins".*
Number of proteins: (Wu et al. 2020) -*"By aligning with the amino acid sequence of SARS PP1ab and analyzing the characteristics of restriction cleavage sites recognized by 3CLpro and PLpro, we speculated 14 specific proteolytic sites of 3CLpro and PLpro in SARS-CoV-2 PP1ab (Fig. 2B). PLpro cleaves three sites at 181–182, 818–819, and 2763–2764 at the N-terminus and 3CLpro cuts at the other 11 sites at the C-terminus, and forming 15 non-structural proteins."*
Evolution rate: (Koyama et al. 2020) - *"Mutation rates estimated for SARS, MERS, and OC43 show a large range, covering a span of 0.27 to 2.38 substitutions ×10-3 / site / year (10-16)."* Recent unpublished evidence also suggest this rate is of the same order of magnitude in SARS-CoV-2.
Mutation rate: (Sanjuan et al. 2010) - *"Murine hepatitis virus … Therefore, the corrected estimate of the mutation rate is $\mu_{s/n/c}$ = 1.9x10$^{-6}$ / 0.55 = 3.5 x 10$^{-6}$."*
Genome similarity: For all species except pangolin: (Wu et al. 2020) - *"After phylogenetic analysis and sequence alignment of 23 coronaviruses from various species. We found three coronaviruses from bat (96%, 88% and 88% for Bat-Cov RaTG13, bat-SL-CoVZXC12 and bat-SL-CoVZC45, respectively) have the highest genome sequence identity to SARS-CoV-2 (Fig. 1A). Moreover, as shown in Fig. 1B, Bat-Cov RaTG13 exhibited the closest linkage with SARS-CoV-2. These phylogenetic evidences suggest that SARS-CoV-2 may be evolved from bat CoVs, especially RaTG13. Among all coronaviruses from human, SARS-CoV (80%) exhibited the highest genome sequence identity to SARS-CoV-2. And MERS/isolate NL13845 also has 50% identity with SARS-CoV-2."* For pangolin: (Zhang et al. 2020) - Figure 3

## Replication Timescales

Virion entry into cell: (Schneider et al. 2012) - *"Previous experiments had revealed that virus is internalized within 15 min"* and (Ng et al. 2003) - *"Within the first 10 min, some virus particles were internalised into vacuoles (arrow) that were just below the plasma membrane surface (Fig. 2, arrows). … The observation at 15 min postinfection (p.i.), did not differ much from 10 min p.i. (Fig. 4a)"*
Eclipse period: (Schneider et al. 2012) - *"SARS-CoV replication cycle from adsorption to release of infectious progeny takes about 7 to 8 h (data not shown)."* and (Harcourt et al. 2020) - Figure 4 shows virions are released after 12-36 hrs but because this is multi-step growth this represents an upper bound for the replication cycle.
Burst size: (Hirano et al. 1976) - *"The average per-cell yield of active virus was estimated to be about 6–7× 10$^2$ plaque-forming units."* This data is for MHV, more research is needed to verify these values for SARS-CoV-2.

## Host Cells

Type: (Shieh et al. 2005) - *"Immunohistochemical and in situ hybridization assays demonstrated evidence of SARS-associated coronavirus (SARS-CoV) infection in various respiratory epithelial cells, predominantly type II pneumocytes, and in alveolar macrophages in the lung."* and (Walls et al. 2020) - *"SARS-CoV-2 uses ACE2 to enter target cells"* and (Rockx et al. 2020) - *"In SARS-CoV-2-infected macaques, virus was excreted from nose and throat in absence of clinical signs, and detected in type I and II pneumocytes in foci of diffuse alveolar damage and mucous glands of the nasal cavity.…In the upper respiratory tract, there was focal 5 or locally extensive SARS-CoV-2 antigen expression in epithelial cells of mucous glands in the nasal cavity (septum or concha) of all four macaques, without any associated histological lesions (fig. 2I)."*
Type I and Type II pneumocyte and alveolar macrophage cell number: (Crapo et al. 1982) - Table 4 and (Stone et al. 1992) - Table 5
Epithelial cells in mucous gland cell number and volume: (ICRP 1975) - surface area of nasal cavity, (Tos & Mogensen, 1976) and (Tos & Mogensen, 1977) - mucous gland density, (Widdicombe 2019) - mucous gland volume, (Ordoñez et al. 2001) and (Mercer et al. 1994) - mucous cell volume. We divide the mucous gland volume by the mucous cell volume to arrive at the total number of mucous cells in a mucous gland. We multiply the surface density of mucous glands by the surface area of the nasal cavity to arrive at the total number of mucous glands, and then multiply the total number of mucous glands by the number of mucous cells per mucous gland.
Type II pneumocyte volume: (Fehrenbach et al. 1995) - *"Morphometry revealed that although inter-individual variation due to some oedematous swelling was present, the cells were in a normal size range as indicated by an estimated mean volume of 763 ± 64 μm$^3$"*
Alveolar macrophage volume: (Crapo et al. 1982) - *"Alveolar macrophages were found to be the largest cell in the populations studied, having a mean volume of 2,491 μm$^3$"*

## Concentration

Nasopharynx, Throat, Stool, and Sputum: (Woelfel et al. 2020) - Figure 2. and (Kim et al. 2020) - Figure 1 and (Pan et al. 2020) - Figure. We took the maximal viral load for each patient in nasopharyngeal swabs, throat swabs, stool or in sputum.

## Antibody Response - Seroconversion

Seroconversion time (time period until a specific antibody becomes detectable in the blood): (Zhao et al. 2020) - *"The seroconversion sequentially appeared for Ab, IgM and then IgG, with a median time of 11, 12 and 14 days, respectively"* and (To et al. 2020) - *"For 16 patients with serum samples available 14 days or longer after symptom onset, rates of seropositivity were 94% for anti-NP IgG (n=15), 88% for anti-NP IgM (n=14), 100% for anti-RBD IgG (n=16), and 94% for anti-RBD IgM (n=15)"*
Maintenance of antibody response to virus: (Wu et al. 2007) - *"Among 176 patients who had had severe acute respiratory syndrome (SARS), SARS-specific antibodies were maintained for an average of 2 years, and significant reduction of immunoglobulin G–positive percentage and titers occurred in the third year."*

## Virus Environmental Stability

Half life on surfaces: (van Doremalen et al. 2020) - For half-lives we use Supplementary Table 1. For time to decay from ~10$^4$ to ~10 TCID$_{50}$/L$^{-1}$ air or mL$^{-1}$ medium, we use the first time titer reached detection limit in Figure 1A for surfaces. For aerosols, we use ten half-life values (1000-fold decrease from 10$^4$ to 10, meaning 10 halvings of concentration) from Supplementary Table 1. More studies are urgently needed to clarify the implications of virion stability on the probability of infection from aerosols or surfaces.
RNA stability on surfaces: (Moriarty et al. 2020) - *"SARS-CoV-2 RNA was identified on a variety of surfaces in cabins of both symptomatic and asymptomatic infected passengers up to 17 days after cabins were vacated on the Diamond Princess but before disinfection procedures had been conducted (Takuya Yamagishi, National Institute of Infectious Diseases, personal communication, 2020)."*

## "Characteristic" Infection Progression in a Single Patient

Basic reproductive number, $R_e$: (Li et al. 2020) - *"Our median estimate of the effective reproductive number, Re—equivalent to the basic reproductive number (R0) at the beginning of the epidemic—is 2.38 (95% CI: 2.04–2.77)"* and (Park et al. 2020) - *"Our estimated R0 from the pooled distribution has a median of 2.9 (95% CI: 2.1–4.5)."*
Latent period (from infection to being able to transmit): (Li et al. 2020) - *"In addition, the median estimates for the latent and infectious periods are approximately 3.69 and 3.48 days, respectively."* and Table 1 and (He et al. 2020) - We use the time it takes the infectiousness to reach half its peak, which happens two days before symptom onset based on Figure 1b. As symptoms arise after 5 days (see incubation period), this means the latent period is about 3 days.
Incubation period (from infection to symptoms): (Lauer et al. 2020) - *"The median incubation period was estimated to be 5.1 days (95% CI, 4.5 to 5.8 days), and 97.5% of those who develop symptoms will do so within 11.5 days (CI, 8.2 to 15.6 days) of infection. These estimates imply that, under conservative assumptions, 101 out of every 10 000 cases (99th percentile, 482) will develop symptoms after 14 days of active monitoring or quarantine."* and (Li et al. 2020) - *"The mean incubation period was 5.2 days (95% confidence interval [CI], 4.1 to 7.0), with the 95th percentile of the distribution at 12.5 days."*
Infectious period (partially overlaps latent period): (Li et al. 2020) - *"In addition, the median estimates for the latent and infectious periods are approximately 3.69 and 3.48 days, respectively."* and Table 1 and (He et al. 2020) - We quantify the interval between half the maximal infectiousness from the infectiousness profile in Figure 1b.
Disease duration: (WHO 2020) - *"Using available preliminary data, the median time from onset to clinical recovery for mild cases is approximately 2 weeks and is 3-6 weeks for patients with severe or critical disease"*
Time until diagnosis: (Xu et al. 2020) - We used data on cases with known symptom onset and case confirmation dates and calculated the median time delay between these two dates.
Case Fatality Rate: (ECDC geographic distribution of cases from 29/03/2020) - We use data from all countries with more than 50 death cases and calculate the uncorrected raw Case Fatality Rate for each country. The range represents the lowest and highest rates observed.
Infected Fatality Rate: (Verity et al. 2020) - *"We obtain an overall IFR estimate for China of 0.66% (0.39%,1.33%)"* and (Ferguson et al. 2020) - *"The IFR estimates from Verity et al.12 have been adjusted to account for a non-uniform attack rate giving an overall IFR of 0.9% (95% credible interval 0.4%-1.4%)."*

## Acknowledgements

We thank Uri Alon, Niv Antonovsky, David Baltimore, Rachel Banks, Arren Bar Even, Naama Barkai, Molly Bassette, Menalu Berihoon, Biana Bernshtein, Pamela Bjorkman, Cecilia Blikstad, Julia Borden, Bill Burkholder, Griffin Chure, Lillian Cohn, Bernadeta Dadonaite, Emmie De wit, Ron Diskin, Ana Duarte, Tal Einav, Avigdor Eldar, Elizabeth Fischer, William Gelbart, Alon Gildoni, Britt Glausinger, Shmuel Gleizer, Dani Gluck, Soichi Hirokawa, Greg Huber, Christina Hueschen, Amit Huppert, Shalev Itzkovitz, Martin Jonikas, Leeat Keren, Gilmor Keshet, Marc Kirschner, Roy Kishony, Amy Kistler, Liad Levi, Sergei Maslov, Adi Millman, Amir Milo, Elad Noor, Gal Ofir, Alan Perelson, Steve Quake, Itai Raveh, Andrew Rennekamp, Tom Roeschinger, Daniel Rokhsar, Alex Rubinsteyn, Gabriel Salmon, Maya Schuldiner, Eran Segal, Ron Sender, Alex Sigal, Maya Shamir, Arik Shams, Mike Springer, Adi Stern, Noam Stern-Ginossar, Lubert Stryer, Dan Tawfik, Boris Veytsman, Aryeh Wides, Tali Wiesel, Anat Yarden, Yossi Yovel, Dudi Zeevi, Mushon Zer Aviv, and Alexander Zlokapa for productive feedback on this manuscript. Figure created using Biorender.